\documentclass[a4paper,11pt]{article}

\usepackage{pos}
\usepackage{siunitx}
\usepackage{setspace}
\usepackage{enumitem}

\usepackage{xcolor}

\usepackage{wrapfig}

\newcommand{\reffig}[1]{Fig.~\ref{#1}}

\newcommand{\refref}[1]{Ref.~\cite{#1}}
\newcommand{\refrefs}[2]{Refs.~\cite{#1}~and~\cite{#2}}

\title{A prototype station of the IceCube-Gen2 Surface Array at the Pierre Auger Observatory}

\ShortTitle{IceCube-Gen2 Surface Prototype at Auger}

\author*[a]{Stef Verpoest}
\author[b]{Carmen Merx}
\author{Benjamin Flaggs$^a$ on behalf of the IceCube-Gen2$^1$ and Pierre Auger$^{c,2}$ Collaborations\\{\normalsize \normalfont(a complete list of authors can be found at the end of the proceedings)}\\}

\affiliation[a]{Bartol Research Institute, Department of Physics and Astronomy, University of Delaware}

\affiliation[b]{Institute for Astroparticle Physics (IAP), Karlsruhe Institute of Technology (KIT),\\76021 Karlsruhe, Germany}

\affiliation[c]{Observatorio Pierre Auger, Av. San Martín Norte 304, 5613 Malargüe, Argentina}

\emailAdd{verpoest@udel.edu}
\emailAdd{analysis@icecube.wisc.edu}
\emailAdd{spokerspersons@auger.org}

\note{Full author list at
\url{https://icecube.wisc.edu/collaboration/authors/}.}
\note{Full author list at \url{https://www.auger.org/archive/authors_icrc_2025.html}.}

\abstract{The detection of extensive air showers using radio antennas has evolved into a mature technique, complementing particle detector arrays by providing sensitivity to the longitudinal development of the showers and enabling an independent determination of the cosmic-ray energy. Both the Pierre Auger Observatory in Argentina and the IceCube Neutrino Observatory at the South Pole have been undergoing upgrades, including the integration of radio antennas. The next-generation neutrino detector IceCube-Gen2 will also feature a surface array for PeV-EeV cosmic-ray detection, consisting of scintillation detectors and radio antennas. Prototype stations for this upgrade have been in operation for several years at both, the South Pole and the Auger Observatory, enabling cross-checks and potentially a cross-calibration of the energy scales between the two experiments. In this contribution, we present an analysis of air showers observed with the radio antennas of the IceCube-Gen2 prototype station, coinciding with detections by the water-Cherenkov detectors of the densest part of the Pierre Auger Observatory’s surface array, featuring a 433 m spacing.}

\ConferenceLogo{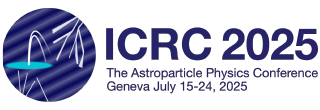}

\FullConference{39th International Cosmic Ray Conference (ICRC2025)\\
 15–24 July 2025\\
Geneva, Switzerland\\}

\begin{document}

\maketitle

\section{Introduction}\label{sec:intro}

The IceCube Neutrino Observatory~\cite{IceCube:2016zyt}, located at the South Pole, includes a surface array (IceTop), consisting of ice-Cherenkov detectors which enable air-shower detection in the PeV to EeV range. The planned extension of the observatory, IceCube-Gen2~\cite{IceCube-Gen2:2020qha}, is also designed to include a surface array~\cite{IceCube-Gen2:2021aek}. The design consists of scintillator detectors rather than Cherenkov detectors for the detection of shower particles. The scintillators will be complemented by antennas for the detection of radio emission produced by the air shower. A prototype detector station comprising eight scintillators and three antennas has been running for several years at the South Pole~\cite{IceCube:2021epf, IceCube:2023rrl}, and two more stations have been deployed in the 2024/2025 Pole season~\cite{SAE:2025icrc}.

\begin{wrapfigure}{r}{0.5\textwidth}
    \centering
    \includegraphics[width=0.48\textwidth]{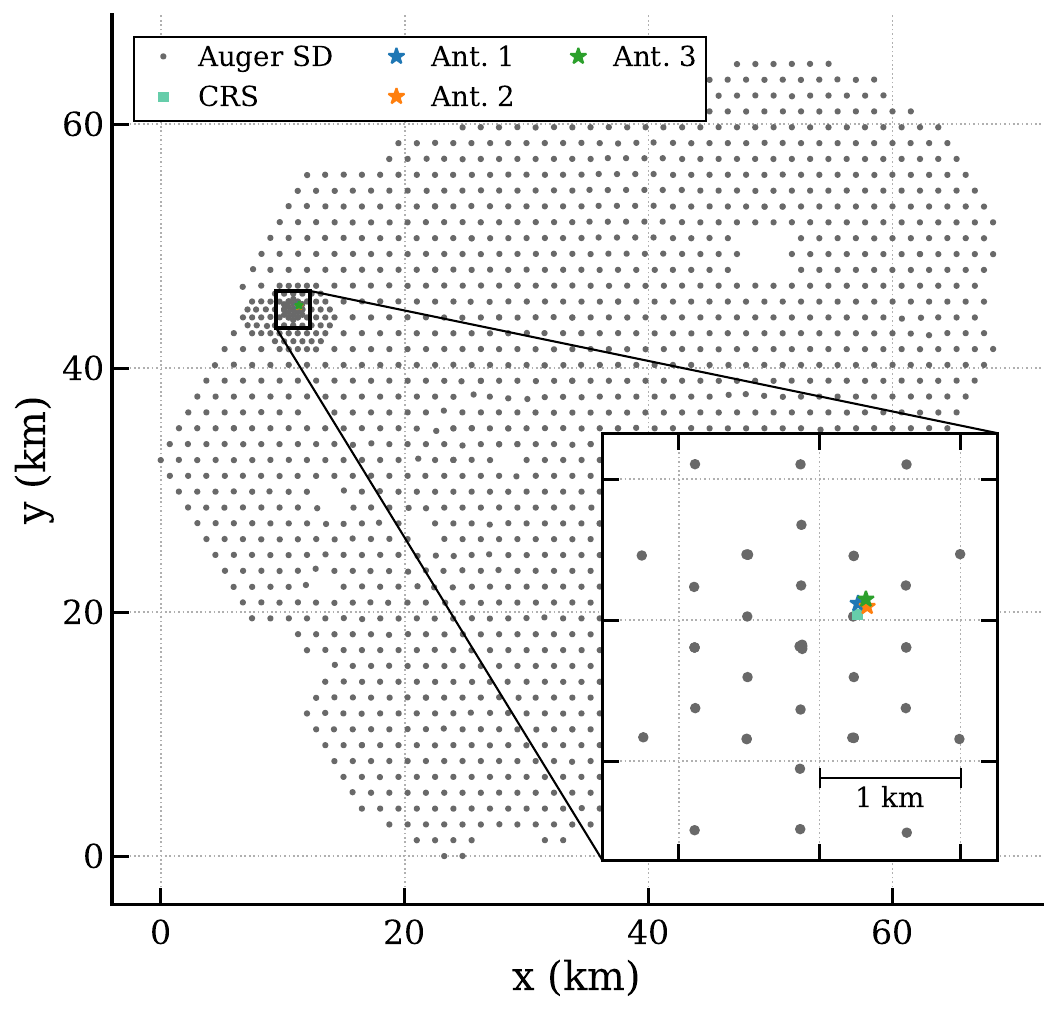}
    \caption{Positions of the SD stations of the Pierre Auger Observatory. The inset shows a zoom of the SD-433 area where the IceCube-Gen2 prototype station is located.}
    \label{fig:Auger}
\end{wrapfigure}

In addition, a prototype station was deployed at the Pierre Auger Observatory~\cite{PierreAuger:2015eyc} in Argentina. The Pierre Auger Observatory is the world's largest cosmic-ray detector, consisting of a surface detector array (SD) of 1660 stations with a spacing of \SI{1500}{\m} overlooked by fluorescence telescopes. The SD also includes two areas of denser instrumentation with \SI{750}{\m} and \SI{433}{\m} spacing. The water-Cherenkov detectors of the SD have recently been upgraded with the addition of a scintillator detector and radio antenna each, as part of the AugerPrime upgrade~\cite{Castellina:2019irv}. The IceCube-Gen2 prototype station is located within the SD-433 area (\reffig{fig:Auger}). It provides an R\&D location that is more accessible than the South Pole and allows exploration of possible synergies between the observatories.

The inclusion of radio antennas in these observatories allows for the detection of the radio emission produced by the charged particles in the shower as a result of the geomagnetic and Askaryan effects. They allow for an independent measurement of the shower energy as well as of the depth of shower maximum with the option of a 100\% duty cycle, and are therefore valuable for calibration purposes, mass composition measurements, and tests of hadronic interaction models.

In this work, we describe the IceCube-Gen2 station deployed at the Pierre Auger Observatory, and demonstrate the successful observation of radio signals from air showers.

\section{Prototype detector station}\label{sec:station}

\begin{figure}
    \centering
    \includegraphics[width=\linewidth]{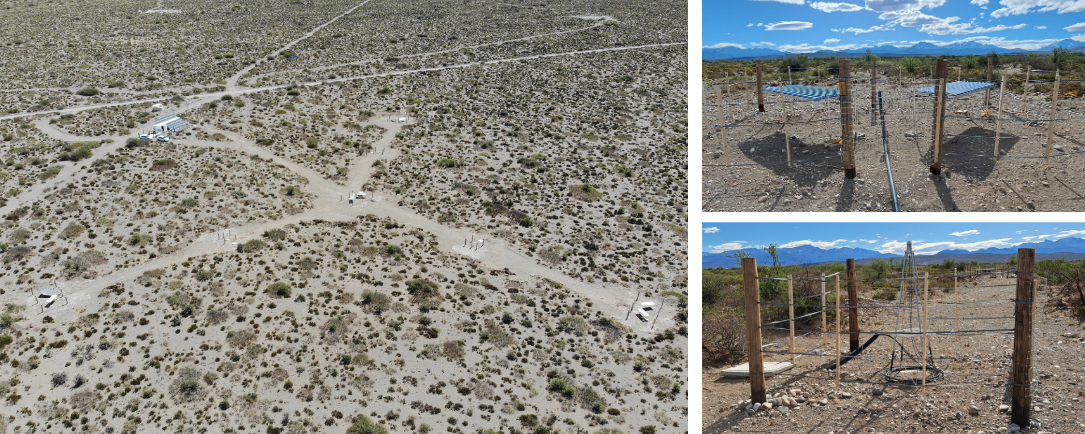}
    \caption{Left: Drone picture of the prototype station at the Pierre Auger Observatory. The container in the top left is the CRS (see text for details). Right: A pair of scintillator detectors (top) and a radio antenna (bottom) of the station.}
    \label{fig:pictures}
\end{figure}

The layout of the detector station deployed at Auger closely follows the reference design of the IceCube-Gen2 station; in-situ pictures are shown in \reffig{fig:pictures}. It consists of a central pair of scintillator detectors from which three arms extend, with another pair of scintillator detectors at the end of each arm. In the middle of each arm, a radio antenna of the SKALAv2 design is located~\cite{skala}. By the central scintillator pair, a box houses the data-acquisition (DAQ) system, known as TAXI\@. The station is located in the SD-433 area of the Observatory, near the \textit{Central Radio Station} (CRS) of the \textit{Auger Engineering Radio Array} (AERA)~\cite{PierreAuger:2012ker}. Power to the station is provided by solar panels and batteries located at the CRS, where also a WhiteRabbit switch is located, which enables communication and provides precision timing information for TAXI\@.

The nominal frequency band of the radio system is \SI{70}{\mega\Hz} to \SI{350}{\mega\Hz}. The radio signals are recorded by DRS4 chips on the TAXI\@. Data have been recorded with sampling rates of \SI{1000}{Msps} and \SI{800}{Msps}, always in 1024 bin waveforms. The readout of the radio data is triggered in two ways: a scintillator trigger, which is formed when six out of eight panels see a signal over threshold within \SI{1}{\micro\second}, and a fixed-rate forced trigger. The former is used to look for air showers, while the latter is used to study the radio background. In this work, the scintillators are only used as a trigger for the radio readout; for more details about their design and operation, see \refref{IceCube:2023pjc}.

\section{Search for radio signals from air showers}\label{sec:analysis}

To demonstrate the detection of radio signals from air showers, a simple analysis is performed, matching scintillator-triggered events with high-amplitude radio signals to events reconstructed by the Auger SD-433 array~\cite{PierreAuger:2023dju}.
We use data recorded between January 2023 and February 2024. During this time, different modes of operation were tested. Importantly, the sampling rate for the radio data was changed from \SI{1000}{Msps} to \SI{800}{Msps} on October 20, 2023. The analysis is performed separately for the two periods, and results are compared for the obtained air-shower samples below. Other changes include the amount of time per day the collection of radio data is running (versus when only scintillator calibration data is collected), the SiPM bias voltage for the scintillators, and the individual scintillator trigger thresholds. This leads to the fractional runtime changing significantly throughout time, and to varying rates of the scintillator-triggered readout of the radio data, spanning a range between about \SI{0.01}{\Hz} and \SI{0.12}{\Hz} throughout the data period. However, in all cases, the energy threshold for the scintillator trigger is expected to be below the energies at which radio emission becomes significant. The fixed-rate trigger, on the other hand, operated at a continuous \SI{0.1}{\Hz} throughout the data period.

In the following, we describe how the event search is performed, after which we discuss the final sample of air showers with identified radio signals.

\subsection{Method}

\begin{wrapfigure}{r}{0.5\textwidth}
    \centering
    \includegraphics[width=0.48\textwidth]{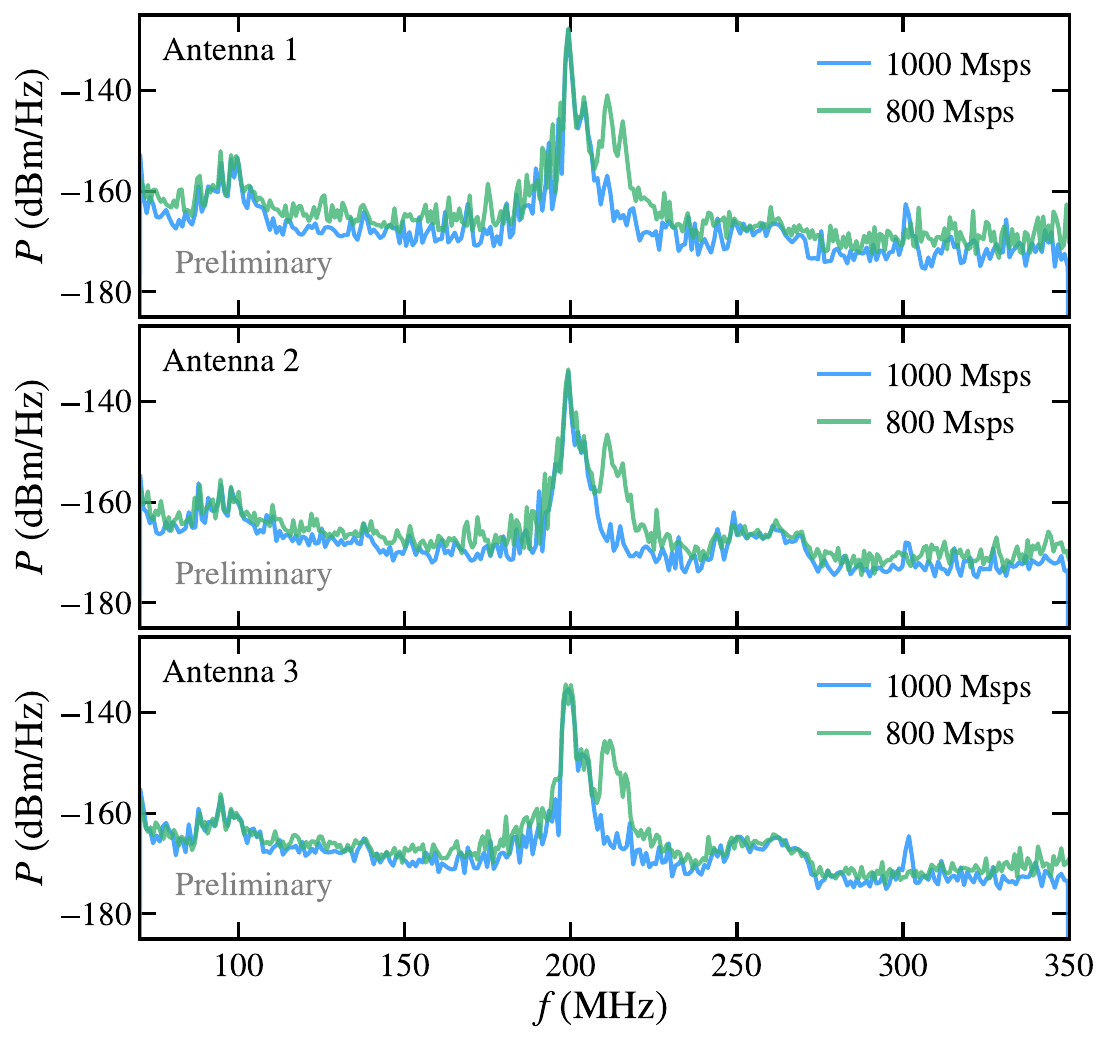}
    \caption{Frequency spectrum of the radio background recorded by the three antennas of the station at the Pierre Auger Observatory. In blue (green), the spectrum obtained in early (late) October 2023 with a sampling rate of \SI{1000}{Msps} (\SI{800}{Msps}).}
    \label{fig:spectrum}
\end{wrapfigure}

The identification of radio signals from air showers is performed in two main steps: 1. the selection of events triggered by the scintillator detectors of the station which also have a large amplitude signal in all three antennas and 2. looking for a corresponding event with a matching trigger time and directional reconstruction in the Auger SD-433 dataset.

The ambient radio spectrum as recorded by the antennas is shown in \reffig{fig:spectrum}. The analysis is performed in the frequency band of \SI{110}{\mega\Hz} to \SI{185}{\mega\Hz} due to the large background at lower and higher frequencies, mainly from FM radio around \SI{100}{\mega\Hz} and from a TV station around \SI{200}{\mega\Hz}. The differences between the spectra recorded at \SI{1000}{Msps} and \SI{800}{Msps} likely reflect a time-dependence in the anthropogenic background.

Details of the analysis steps have been described before in \refref{Verpoest:2024fi}; we summarize them below. For details of the SD-433 event reconstruction, we refer to \refrefs{PierreAuger:2021tmd}{PierreAuger:2023dju}.

\subsubsection*{Selection of candidate radio events}

\begin{itemize}
    \item A basic processing is applied to the radio data, including corrections for artifacts in the waveforms, bandpass filtering to the \SI{110}{\mega\Hz} to \SI{185}{\mega\Hz} range, and the application of a frequency weighting scheme to further reduce the influence of narrow-band background~\cite{IceCube:2021qnf}.

    \item The forced-trigger data is used to obtain distributions of the signal-to-noise ratio (SNR) in radio background waveforms. The 95th percentile of the SNR values is calculated for the next step in the analysis. This is done for each polarization of each antenna separately.

    \item To select candidates for air-shower events with observed radio emission, the scintillator-triggered events are considered. All events which do not have at least one polarization per antenna with an SNR value higher than 95\% of the background SNR values are rejected.

    \item For each passing event, a simple directional reconstruction is applied by fitting a plane shower front to the signal peak times in each antenna.
\end{itemize}

\subsubsection*{Matching with Auger SD-433 events}

\begin{itemize}
    \item The time offset between the IceCube-Gen2 prototype station and Auger SD is determined statistically by comparing a large number of scintillator-triggered events of the station to air-shower events recorded by SD. This is repeated every 12 hours due to a drift between the absolute times of the systems, which is taken into account in further analysis.

    \item For each radio event candidate, SD-433 events in a window of $\pm \SI{50}{\milli\s}$ are considered after correcting for the time difference between the systems. (Due to the time drift between the systems and the imperfect corrections, this window is rather large. However, as the trigger rate of the SD-433 dataset is only around \SI{0.01}{\Hz}, chance coincidences are rare.)
    
    \item For each event found in the time window around the station trigger, the reconstructed directions from the radio signals and the SD-433 reconstruction are compared. To further ensure that the two events belong to the same air shower, the opening angle is required to be less than 5 degrees. (The event matching based on time and opening angle is visualized in \reffig{fig:selection}, which is discussed more in the following section.)

    \item For each selected event, air-shower simulations are performed with CORSIKA~\cite{Heck:1998vt}, using the shower core position, direction, and energy from the SD-433 reconstruction as input parameters, assuming different primaries. The radio emission is simulated with CoREAS~\cite{Huege:2013vt}, and the response of the antennas and electronics chain is simulated. The simulated and measured waveforms are then compared for validation. A template matching is performed to find the best alignment between the two, after which a visual inspection is performed.
\end{itemize}

\begin{figure}
    \centering
    \includegraphics[width=0.39\linewidth]{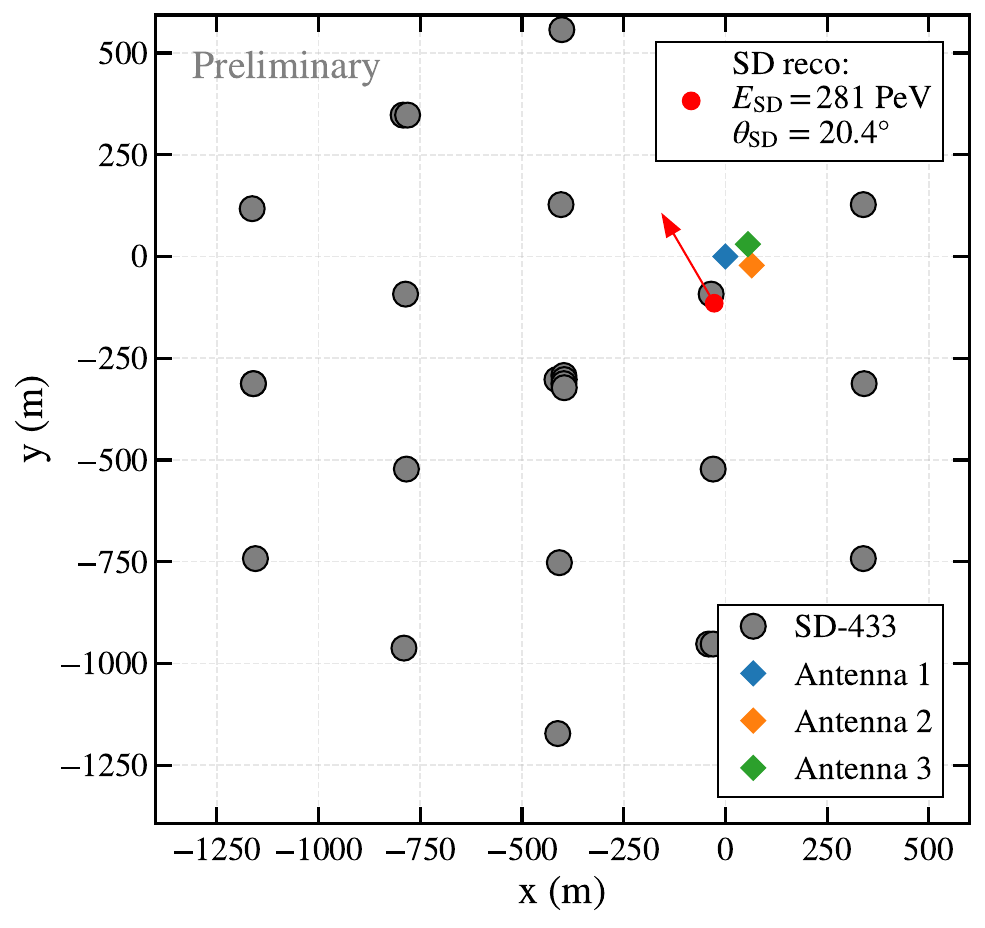}\hfill\includegraphics[width=0.555\linewidth]{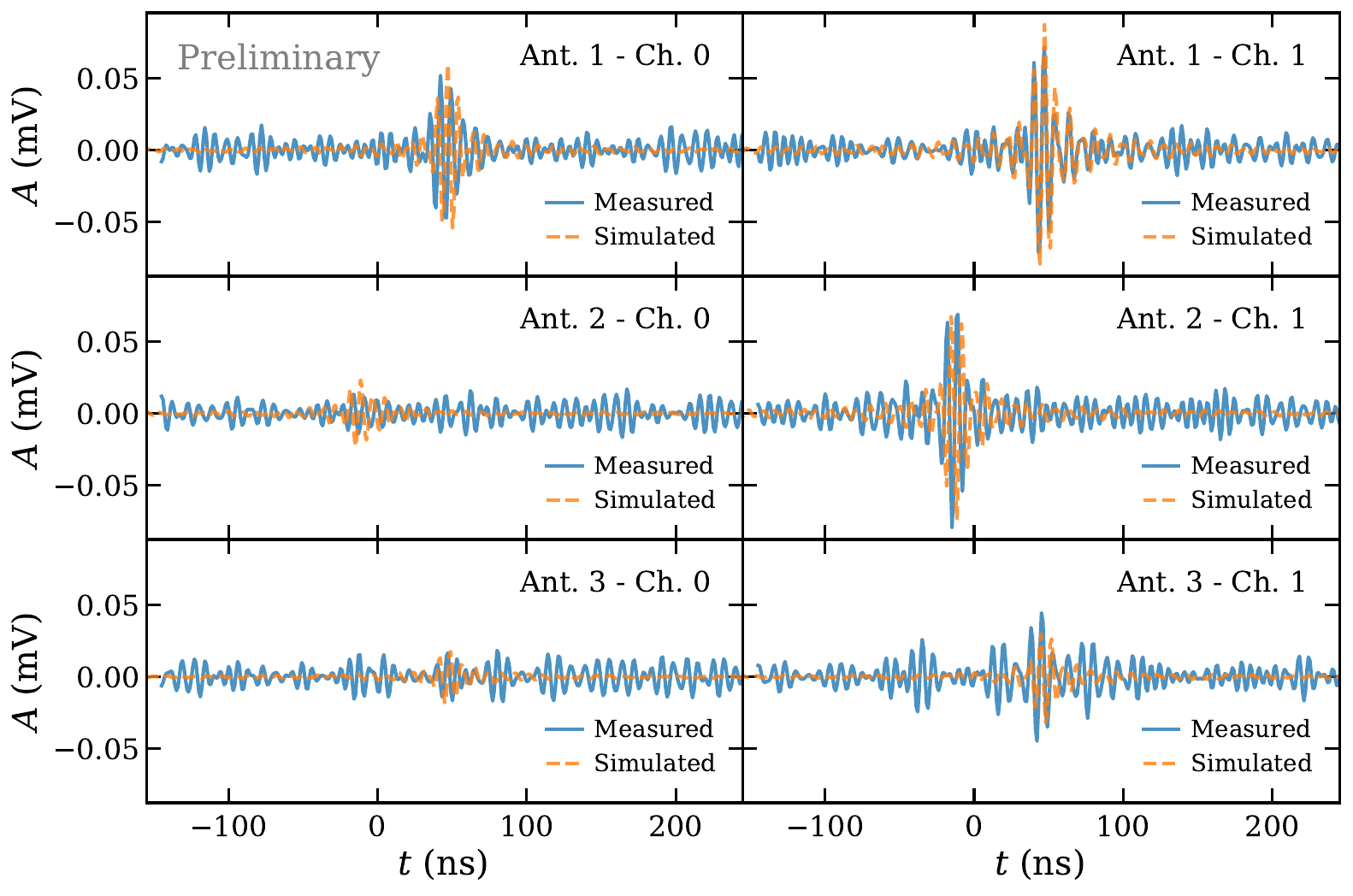}
    \caption{Example of an identified air-shower event. Left: The core position and direction from the SD-433 reconstructions compared to the antenna positions. Right: Comparison of the measured waveforms and waveforms from an air shower simulated based on the SD-433 reconstruction.}
    \label{fig:waveform}
\end{figure}

An example event identified following these steps is shown in \reffig{fig:waveform}. Note that deviations between the simulated and measured waveforms are expected due to uncertainties in the reconstructed core position, direction, and energy, in addition to the unknown depth of shower maximum.


\subsection{Identified air showers}

\begin{figure}
    \centering
    \includegraphics[width=0.5\linewidth]{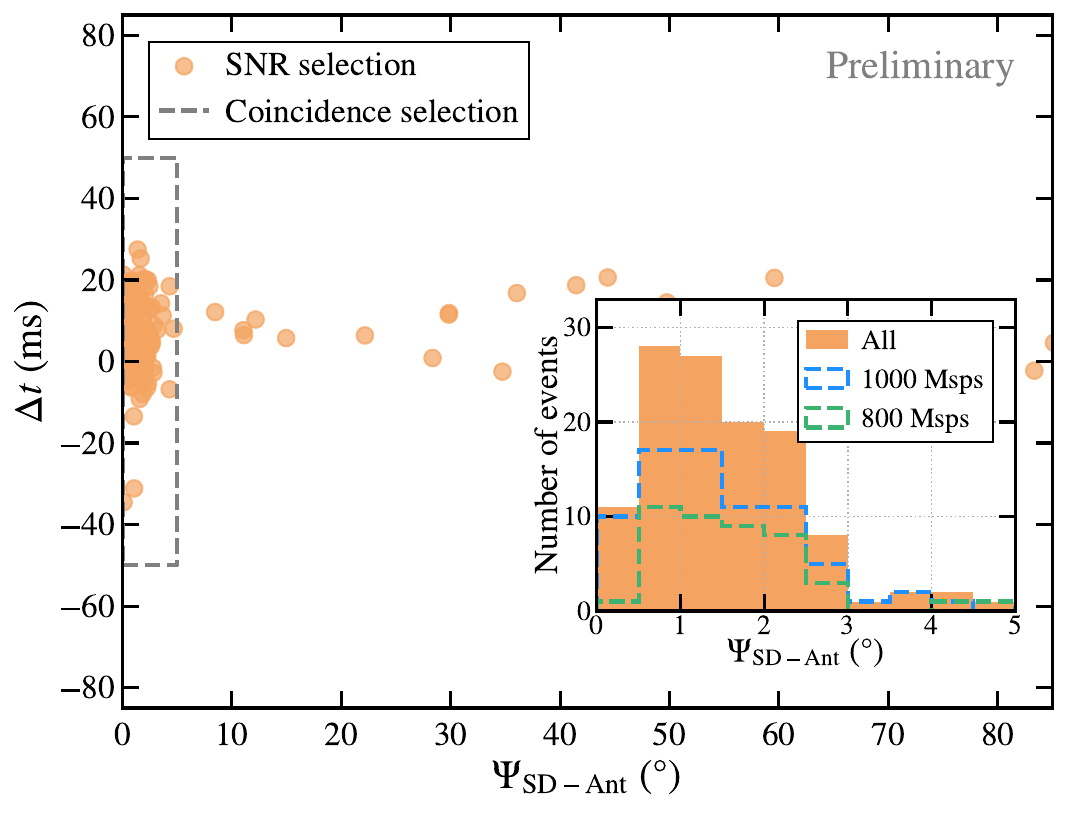}
    \caption{Time difference and opening angle between radio candidate events and SD-433 events. The box defines the selection of coincident events. The inset shows the opening angle distribution of selected events for the total sample, as well as the \SI{1000}{Msps} and \SI{800}{Msps} subsamples.}
    \label{fig:selection}
\end{figure}

We analyzed data recorded between January 1, 2023 and February 12, 2024. The radio data acquisition typically ran for 18 hours per day, with the other hours used to collect scintillator calibration data. Later in 2023, the radio data recording was increased to periods of 21-23 hours per day. Taking this into account, and removing periods of downtime due to hardware and software issues, a total runtime of 241.2 days was available for analysis. Following the method described above, a total of 135 events with radio emission matching an event reconstructed in SD-433 was found. Of these, 17 did not have a successful energy reconstruction from SD-433 and are excluded for the following discussion. After this, the average rate is $0.49 \pm 0.05$ events per day of runtime.

The sampling rate of the radio DAQ was changed from \SI{1000}{Msps} to \SI{800}{Msps} on October 20, 2023. The \SI{1000}{Msps} data corresponds to a runtime of 143.3 days in which 74 events were found. The \SI{800}{Msps} data corresponds to 97.9 days of runtime with 44 identified events. There is no significant difference in the event rates between the two periods.

The identification of the events based on the time and direction observed in both the prototype station and SD-433 is visualized in \reffig{fig:selection} for the full data period included in this analysis. SD-433 events with small time differences compared to the prototype station typically also cluster at small opening angles, indicating that these are real hybrid events. The opening angle distribution of the identified events, shown in the inset of the figure, peaks near $1^\circ$, indicating that even with three radio antennas, a good angular resolution is possible. Note that there is a tail of events at small time differences which have large opening angles; these are likely misreconstructed events which may be recoverable in future work.

\begin{figure}
    \centering
    \includegraphics[width=0.47\linewidth]{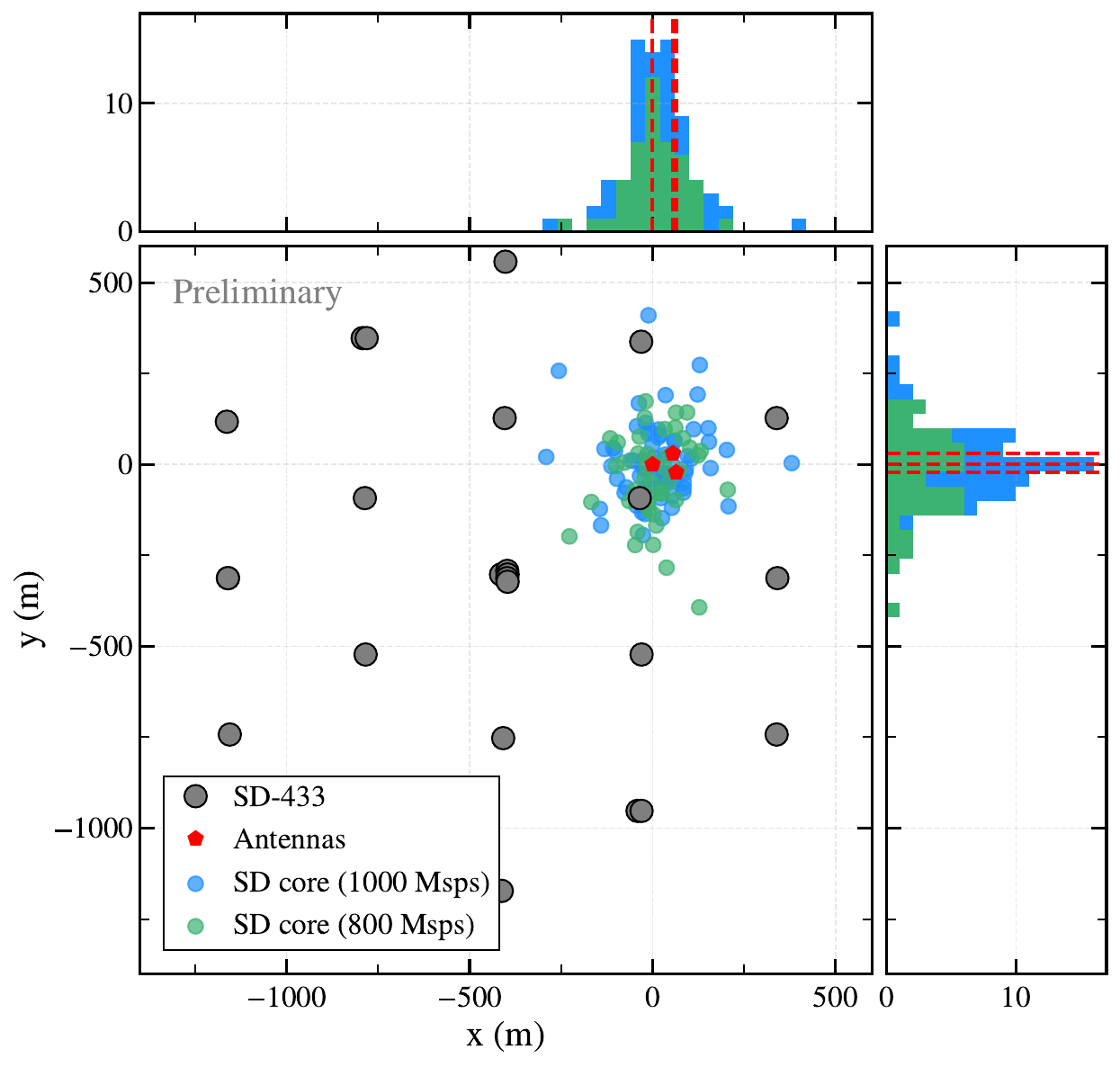}\hfill\includegraphics[width=0.47\linewidth]{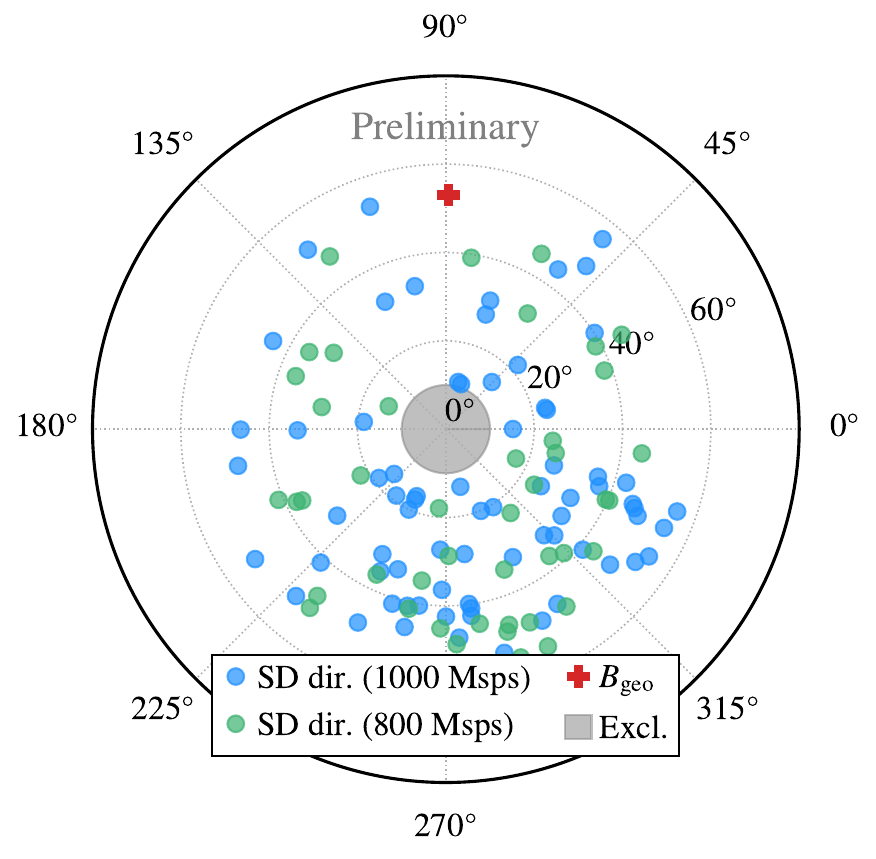}
    \caption{Shower geometry of the identified events as reconstructed with SD-433, for the \SI{1000}{Msps} and \SI{800}{Msps} subsamples. Left: Core positions compared to the position of the antennas. The dashed lines in the histograms represent the antenna positions. The coordinates have been centered on the position of Antenna 1. Right: Reconstructed arrival directions compared to the geomagnetic field direction. (The zenith is excluded due to occasional artifacts in the radio data.)}
    \label{fig:cores}
\end{figure}

The core positions for the identified events, as reconstructed by SD-433, are shown in \reffig{fig:cores} (left). All events cluster near the location of the antennas, mostly within a radius of $\sim \SI{250}{\m}$ around the station center. In \reffig{fig:cores} (right), the reconstructed arrival directions are shown. We observe a tendency for events to appear more frequently at larger angles to the geomagnetic field direction, which is expected due to the radio emission produced by the geomagnetic effect. The area around the zenith is currently excluded from the event selection, as some DAQ artifacts not caught by our cleaning algorithms tend to mimic vertical air showers.

\begin{figure}
    \centering
    \includegraphics[width=0.5\linewidth]{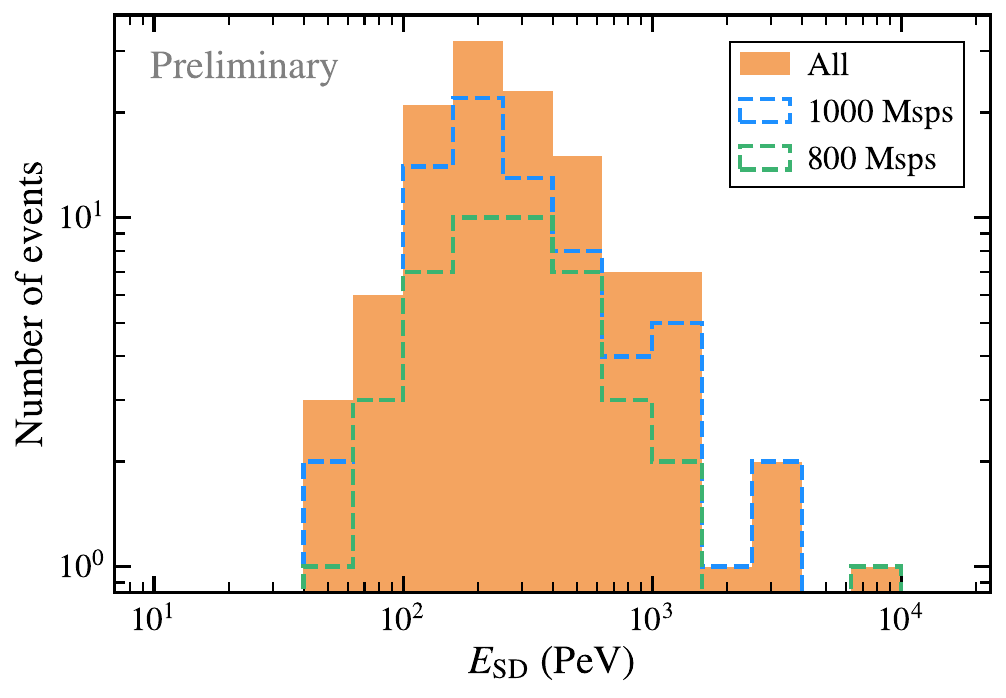}
    \caption{Distribution of the SD-433 reconstructed shower energies, for the total sample as well as the \SI{1000}{Msps} and \SI{800}{Msps} subsamples.}
    \label{fig:energy}
\end{figure}

In \reffig{fig:energy}, the distribution of energies reconstructed by SD-433 is shown. The lowest energy events are around several 10s of PeV and the distribution peaks around \SI{200}{\peta\eV}. SD-433 reaches full efficiency for showers below $40^\circ$ around \SI{63}{\peta\eV}~\cite{PierreAuger:2023dju}. For the future, it would be of interest to study the efficiency for radio observation as a function of the SD-433 energy, given a certain shower geometry. A recently deployed array of 18 SKALA antennas is expected to reach full efficiency for the reconstruction based on radio signals around \SI{300}{\peta\eV} for showers with zenith angles below $30^\circ$~\cite{Verpoest:2025fyj}.

\section{Summary \& Outlook}\label{sec:conclusion}

A prototype detector station for the surface component of the IceCube-Gen2 Observatory, consisting of scintillator panels and radio antennas, was deployed at the Pierre Auger Observatory and has been taking data for several years. First data from the station was used to successfully observe the radio emission from air showers trigger by the accompanying scintillators. By matching with showers reconstructed using the Auger SD-433 array, 118 events were identified in about 241 days of runtime, corresponding to a rate of $0.49\pm0.05$ events per day.

The selection of events with radio signals could be improved in the future, e.g. through more sophisticated SNR-based cuts, the use of neural networks for classifying and/or denoising~\cite{Schroeder:2024bzf}, or by a more focused search for radio signals for events already reconstructed with SD-433. The successful radio detection of air showers with the station at this location suggests that future combined studies with the surface array of IceCube(-Gen2) and the Pierre Auger Observatory could be explored. Using the same antennas at both sites will reduce systematic uncertainties, e.g. in a comparison of the cosmic-ray energy scales in the overlapping energy range of the observatories~\cite{Coleman:2022abf}.

\bibliographystyle{ICRC}
\bibliography{main}

\providecommand{\href}[2]{#2}\begingroup\raggedright\setstretch{0.01}\begin{thebibliography}{10}

\bibitem{IceCube:2016zyt}
{\bfseries IceCube} Collaboration, M.~G. Aartsen {\em et~al.}, \href{http://dx.doi.org/10.1088/1748-0221/12/03/P03012}{{\em JINST} {\bfseries 12} no.~03, (2017) P03012}. [Erratum: JINST 19, E05001 (2024)].

\bibitem{IceCube-Gen2:2020qha}
{\bfseries IceCube-Gen2} Collaboration, M.~G. Aartsen {\em et~al.}, \href{http://dx.doi.org/10.1088/1361-6471/abbd48}{{\em J. Phys. G} {\bfseries 48} no.~6, (2021) 060501}.

\bibitem{IceCube-Gen2:2021aek}
{\bfseries IceCube-Gen2} Collaboration, F.~G. Schroeder {\em et~al.}, \href{http://dx.doi.org/10.22323/1.395.0407}{{\em PoS} {\bfseries ICRC2021} (2021) 407}.

\bibitem{IceCube:2021epf}
{\bfseries IceCube} Collaboration, R.~Abbasi {\em et~al.}, \href{http://dx.doi.org/10.22323/1.395.0314}{{\em PoS} {\bfseries ICRC2021} (2021) 314}.

\bibitem{IceCube:2023rrl}
{\bfseries IceCube} Collaboration, A.~Rehman {\em et~al.}, \href{http://dx.doi.org/10.22323/1.444.0291}{{\em PoS} {\bfseries ICRC2023} (2023) 291}.

\bibitem{SAE:2025icrc}
{\bfseries IceCube} Collaboration, M.~Venugopal, {\em PoS} {\bfseries ICRC2025} (these proceedings) 427.

\bibitem{PierreAuger:2015eyc}
{\bfseries Pierre Auger} Collaboration, A.~Aab {\em et~al.}, \href{http://dx.doi.org/10.1016/j.nima.2015.06.058}{{\em Nucl. Instrum. Meth. A} {\bfseries 798} (2015) 172--213}.

\bibitem{Castellina:2019irv}
{\bfseries Pierre Auger} Collaboration, A.~Castellina, \href{http://dx.doi.org/10.1051/epjconf/201921006002}{{\em EPJ Web Conf.} {\bfseries 210} (2019) 06002}.

\bibitem{skala}
E.~{de Lera Acedo}, N.~{Razavi-Ghods}, N.~{Troop}, N.~{Drought}, and A.~J. {Faulkner}, \href{http://dx.doi.org/10.1007/s10686-015-9439-0}{{\em Experimental Astronomy} {\bfseries 39} no.~3, (Oct., 2015) 567--594}.

\bibitem{PierreAuger:2012ker}
{\bfseries Pierre Auger} Collaboration, P.~Abreu {\em et~al.}, \href{http://dx.doi.org/10.1088/1748-0221/7/10/P10011}{{\em JINST} {\bfseries 7} (2012) P10011}.

\bibitem{IceCube:2023pjc}
{\bfseries IceCube} Collaboration, F.~G. Schroeder {\em et~al.}, \href{http://dx.doi.org/10.22323/1.444.0342}{{\em PoS} {\bfseries ICRC2023} (2023) 342}.

\bibitem{PierreAuger:2023dju}
{\bfseries Pierre Auger} Collaboration, G.~Brichetto~Orquera {\em et~al.}, \href{http://dx.doi.org/10.22323/1.444.0398}{{\em PoS} {\bfseries ICRC2023} (2023) 398}.

\bibitem{Verpoest:2024fi}
{\textbf{IceCube-Gen2} and \textbf{Pierre Auger} Collaborations, S. Verpoest \textit{et al.}}, \href{http://dx.doi.org/10.22323/1.470.0037}{{\em PoS} {\bfseries ARENA2024} (2024) 037}.

\bibitem{PierreAuger:2021tmd}
{\bfseries Pierre Auger} Collaboration, P.~Abreu {\em et~al.}, \href{http://dx.doi.org/10.22323/1.395.0224}{{\em PoS} {\bfseries ICRC2021} (2021) 224}.

\bibitem{IceCube:2021qnf}
{\bfseries IceCube} Collaboration, R.~Abbasi {\em et~al.}, \href{http://dx.doi.org/10.22323/1.395.0317}{{\em PoS} {\bfseries ICRC2021} (2021) 317}.

\bibitem{Heck:1998vt}
D.~Heck {\em et~al.}, {\em Forschungszentrum Karlsruhe Report FZKA-6019} (1998) .

\bibitem{Huege:2013vt}
T.~Huege, M.~Ludwig, and C.~W. James, \href{http://dx.doi.org/10.1063/1.4807534}{{\em AIP Conf. Proc.} {\bfseries 1535} no.~1, (2013) 128}.

\bibitem{Verpoest:2025fyj}
S.~Verpoest, F.~Schroeder, A.~Novikov, A.~Coleman, B.~Flaggs, A.~Weindl, M.~Venugopal, C.~Merx, and A.~Haungs, \href{http://dx.doi.org/10.22323/1.484.0122}{{\em PoS} {\bfseries UHECR2024} (2025) 122}.

\bibitem{Schroeder:2024bzf}
{\bfseries IceCube} Collaboration, F.~G. Schroeder and A.~Rehman, \href{http://dx.doi.org/10.22323/1.470.0034}{{\em PoS} {\bfseries ARENA2024} (2024) 034}.

\bibitem{Coleman:2022abf}
A.~Coleman {\em et~al.}, \href{http://dx.doi.org/10.1016/j.astropartphys.2023.102819}{{\em Astropart. Phys.} {\bfseries 149} (2023) 102819}.

\end{thebibliography}\endgroup

\clearpage

\section*{Full Author List: IceCube-Gen2 Collaboration}

\scriptsize
\noindent
R. Abbasi$^{16}$,
M. Ackermann$^{76}$,
J. Adams$^{21}$,
S. K. Agarwalla$^{46,\: {\rm a}}$,
J. A. Aguilar$^{10}$,
M. Ahlers$^{25}$,
J.M. Alameddine$^{26}$,
S. Ali$^{39}$,
N. M. Amin$^{52}$,
K. Andeen$^{49}$
% ...
\\
\\
% ...
$^{10}$ Universit{\'e} Libre de Bruxelles, Science Faculty CP230, B-1050 Brussels, Belgium \\
% ...
$^{16}$ Department of Physics, Loyola University Chicago, Chicago, IL 60660, USA \\
% ...
$^{21}$ Dept. of Physics and Astronomy, University of Canterbury, Private Bag 4800, Christchurch, New Zealand \\
% ...
$^{25}$ Niels Bohr Institute, University of Copenhagen, DK-2100 Copenhagen, Denmark \\
$^{26}$ Dept. of Physics, TU Dortmund University, D-44221 Dortmund, Germany \\
% ...
$^{39}$ Dept. of Physics and Astronomy, University of Kansas, Lawrence, KS 66045, USA \\
% ...
$^{46}$ Dept. of Physics and Wisconsin IceCube Particle Astrophysics Center, University of Wisconsin{\textemdash}Madison, Madison, WI 53706, USA \\
% ...
$^{49}$ Department of Physics, Marquette University, Milwaukee, WI 53201, USA \\
% ...
$^{52}$ Bartol Research Institute and Dept. of Physics and Astronomy, University of Delaware, Newark, DE 19716, USA \\
% ...
$^{76}$ Deutsches Elektronen-Synchrotron DESY, Platanenallee 6, D-15738 Zeuthen, Germany \\
$^{\rm a}$ also at Institute of Physics, Sachivalaya Marg, Sainik School Post, Bhubaneswar 751005, India \\
% ...

\subsection*{Acknowledgments}

\noindent
The authors gratefully acknowledge the support from the following agencies and institutions:
USA {\textendash} U.S. National Science Foundation-Office of Polar Programs,
U.S. National Science Foundation-Physics Division,
U.S. National Science Foundation-EPSCoR,
U.S. National Science Foundation-Office of Advanced Cyberinfrastructure,
Wisconsin Alumni Research Foundation,
Center for High Throughput Computing (CHTC) at the University of Wisconsin{\textendash}Madison,
Open Science Grid (OSG),
Partnership to Advance Throughput Computing (PATh),
Advanced Cyberinfrastructure Coordination Ecosystem: Services {\&} Support (ACCESS),
Frontera and Ranch computing project at the Texas Advanced Computing Center,
U.S. Department of Energy-National Energy Research Scientific Computing Center,
Particle astrophysics research computing center at the University of Maryland,
Institute for Cyber-Enabled Research at Michigan State University,
Astroparticle physics computational facility at Marquette University,
NVIDIA Corporation,
and Google Cloud Platform;
Belgium {\textendash} Funds for Scientific Research (FRS-FNRS and FWO),
FWO Odysseus and Big Science programmes,
and Belgian Federal Science Policy Office (Belspo);
Germany {\textendash} Bundesministerium f{\"u}r Forschung, Technologie und Raumfahrt (BMFTR),
Deutsche Forschungsgemeinschaft (DFG),
Helmholtz Alliance for Astroparticle Physics (HAP),
Initiative and Networking Fund of the Helmholtz Association,
Deutsches Elektronen Synchrotron (DESY),
and High Performance Computing cluster of the RWTH Aachen;
Sweden {\textendash} Swedish Research Council,
Swedish Polar Research Secretariat,
Swedish National Infrastructure for Computing (SNIC),
and Knut and Alice Wallenberg Foundation;
European Union {\textendash} EGI Advanced Computing for research;
Australia {\textendash} Australian Research Council;
Canada {\textendash} Natural Sciences and Engineering Research Council of Canada,
Calcul Qu{\'e}bec, Compute Ontario, Canada Foundation for Innovation, WestGrid, and Digital Research Alliance of Canada;
Denmark {\textendash} Villum Fonden, Carlsberg Foundation, and European Commission;
New Zealand {\textendash} Marsden Fund;
Japan {\textendash} Japan Society for Promotion of Science (JSPS)
and Institute for Global Prominent Research (IGPR) of Chiba University;
Korea {\textendash} National Research Foundation of Korea (NRF);
Switzerland {\textendash} Swiss National Science Foundation (SNSF).
\clearpage
\section*{Full Author List: Pierre Auger Collaboration}

A.~Abdul Halim$^{13}$,
P.~Abreu$^{70}$,
M.~Aglietta$^{53,51}$,
I.~Allekotte$^{1}$,
K.~Almeida Cheminant$^{78,77}$,
A.~Almela$^{7,12}$,
R.~Aloisio$^{44,45}$,
J.~Alvarez-Mu\~niz$^{76}$,
A.~Ambrosone$^{44}$,
J.~Ammerman Yebra$^{76}$
% ...

\begin{description}[labelsep=0.2em,align=right,labelwidth=0.7em,labelindent=0em,leftmargin=2em,noitemsep,before={\renewcommand\makelabel[1]{##1 }}]
\item[$^{1}$] Centro At\'omico Bariloche and Instituto Balseiro (CNEA-UNCuyo-CONICET), San Carlos de Bariloche, Argentina
\item[$^{2}$] Departamento de F\'\i{}sica and Departamento de Ciencias de la Atm\'osfera y los Oc\'eanos, FCEyN, Universidad de Buenos Aires and CONICET, Buenos Aires, Argentina
\item[$^{3}$] IFLP, Universidad Nacional de La Plata and CONICET, La Plata, Argentina
\item[$^{4}$] Instituto de Astronom\'\i{}a y F\'\i{}sica del Espacio (IAFE, CONICET-UBA), Buenos Aires, Argentina
\item[$^{5}$] Instituto de F\'\i{}sica de Rosario (IFIR) -- CONICET/U.N.R.\ and Facultad de Ciencias Bioqu\'\i{}micas y Farmac\'euticas U.N.R., Rosario, Argentina
\item[$^{6}$] Instituto de Tecnolog\'\i{}as en Detecci\'on y Astropart\'\i{}culas (CNEA, CONICET, UNSAM), and Universidad Tecnol\'ogica Nacional -- Facultad Regional Mendoza (CONICET/CNEA), Mendoza, Argentina
\item[$^{7}$] Instituto de Tecnolog\'\i{}as en Detecci\'on y Astropart\'\i{}culas (CNEA, CONICET, UNSAM), Buenos Aires, Argentina
\item[$^{8}$] International Center of Advanced Studies and Instituto de Ciencias F\'\i{}sicas, ECyT-UNSAM and CONICET, Campus Miguelete -- San Mart\'\i{}n, Buenos Aires, Argentina
\item[$^{9}$] Laboratorio Atm\'osfera -- Departamento de Investigaciones en L\'aseres y sus Aplicaciones -- UNIDEF (CITEDEF-CONICET), Argentina
\item[$^{10}$] Observatorio Pierre Auger, Malarg\"ue, Argentina
\item[$^{11}$] Observatorio Pierre Auger and Comisi\'on Nacional de Energ\'\i{}a At\'omica, Malarg\"ue, Argentina
\item[$^{12}$] Universidad Tecnol\'ogica Nacional -- Facultad Regional Buenos Aires, Buenos Aires, Argentina
\item[$^{13}$] University of Adelaide, Adelaide, S.A., Australia
\item[$^{14}$] Universit\'e Libre de Bruxelles (ULB), Brussels, Belgium
\item[$^{15}$] Vrije Universiteit Brussels, Brussels, Belgium
\item[$^{16}$] Centro Brasileiro de Pesquisas Fisicas, Rio de Janeiro, RJ, Brazil
\item[$^{17}$] Centro Federal de Educa\c{c}\~ao Tecnol\'ogica Celso Suckow da Fonseca, Petropolis, Brazil
\item[$^{18}$] Instituto Federal de Educa\c{c}\~ao, Ci\^encia e Tecnologia do Rio de Janeiro (IFRJ), Brazil
\item[$^{19}$] Universidade de S\~ao Paulo, Escola de Engenharia de Lorena, Lorena, SP, Brazil
\item[$^{20}$] Universidade de S\~ao Paulo, Instituto de F\'\i{}sica de S\~ao Carlos, S\~ao Carlos, SP, Brazil
\item[$^{21}$] Universidade de S\~ao Paulo, Instituto de F\'\i{}sica, S\~ao Paulo, SP, Brazil
\item[$^{22}$] Universidade Estadual de Campinas (UNICAMP), IFGW, Campinas, SP, Brazil
\item[$^{23}$] Universidade Estadual de Feira de Santana, Feira de Santana, Brazil
\item[$^{24}$] Universidade Federal de Campina Grande, Centro de Ciencias e Tecnologia, Campina Grande, Brazil
\item[$^{25}$] Universidade Federal do ABC, Santo Andr\'e, SP, Brazil
\item[$^{26}$] Universidade Federal do Paran\'a, Setor Palotina, Palotina, Brazil
\item[$^{27}$] Universidade Federal do Rio de Janeiro, Instituto de F\'\i{}sica, Rio de Janeiro, RJ, Brazil
\item[$^{28}$] Universidad de Medell\'\i{}n, Medell\'\i{}n, Colombia
\item[$^{29}$] Universidad Industrial de Santander, Bucaramanga, Colombia
\item[$^{30}$] Charles University, Faculty of Mathematics and Physics, Institute of Particle and Nuclear Physics, Prague, Czech Republic
\item[$^{31}$] Institute of Physics of the Czech Academy of Sciences, Prague, Czech Republic
\item[$^{32}$] Palacky University, Olomouc, Czech Republic
\item[$^{33}$] CNRS/IN2P3, IJCLab, Universit\'e Paris-Saclay, Orsay, France
\item[$^{34}$] Laboratoire de Physique Nucl\'eaire et de Hautes Energies (LPNHE), Sorbonne Universit\'e, Universit\'e de Paris, CNRS-IN2P3, Paris, France
\item[$^{35}$] Univ.\ Grenoble Alpes, CNRS, Grenoble Institute of Engineering Univ.\ Grenoble Alpes, LPSC-IN2P3, 38000 Grenoble, France
\item[$^{36}$] Universit\'e Paris-Saclay, CNRS/IN2P3, IJCLab, Orsay, France
\item[$^{37}$] Bergische Universit\"at Wuppertal, Department of Physics, Wuppertal, Germany
\item[$^{38}$] Karlsruhe Institute of Technology (KIT), Institute for Experimental Particle Physics, Karlsruhe, Germany
\item[$^{39}$] Karlsruhe Institute of Technology (KIT), Institut f\"ur Prozessdatenverarbeitung und Elektronik, Karlsruhe, Germany
\item[$^{40}$] Karlsruhe Institute of Technology (KIT), Institute for Astroparticle Physics, Karlsruhe, Germany
\item[$^{41}$] RWTH Aachen University, III.\ Physikalisches Institut A, Aachen, Germany
\item[$^{42}$] Universit\"at Hamburg, II.\ Institut f\"ur Theoretische Physik, Hamburg, Germany
\item[$^{43}$] Universit\"at Siegen, Department Physik -- Experimentelle Teilchenphysik, Siegen, Germany
\item[$^{44}$] Gran Sasso Science Institute, L'Aquila, Italy
\item[$^{45}$] INFN Laboratori Nazionali del Gran Sasso, Assergi (L'Aquila), Italy
\item[$^{46}$] INFN, Sezione di Catania, Catania, Italy
\item[$^{47}$] INFN, Sezione di Lecce, Lecce, Italy
\item[$^{48}$] INFN, Sezione di Milano, Milano, Italy
\item[$^{49}$] INFN, Sezione di Napoli, Napoli, Italy
\item[$^{50}$] INFN, Sezione di Roma ``Tor Vergata'', Roma, Italy
\item[$^{51}$] INFN, Sezione di Torino, Torino, Italy
\item[$^{52}$] Istituto di Astrofisica Spaziale e Fisica Cosmica di Palermo (INAF), Palermo, Italy
\item[$^{53}$] Osservatorio Astrofisico di Torino (INAF), Torino, Italy
\item[$^{54}$] Politecnico di Milano, Dipartimento di Scienze e Tecnologie Aerospaziali , Milano, Italy
\item[$^{55}$] Universit\`a del Salento, Dipartimento di Matematica e Fisica ``E.\ De Giorgi'', Lecce, Italy
\item[$^{56}$] Universit\`a dell'Aquila, Dipartimento di Scienze Fisiche e Chimiche, L'Aquila, Italy
\item[$^{57}$] Universit\`a di Catania, Dipartimento di Fisica e Astronomia ``Ettore Majorana``, Catania, Italy
\item[$^{58}$] Universit\`a di Milano, Dipartimento di Fisica, Milano, Italy
\item[$^{59}$] Universit\`a di Napoli ``Federico II'', Dipartimento di Fisica ``Ettore Pancini'', Napoli, Italy
\item[$^{60}$] Universit\`a di Palermo, Dipartimento di Fisica e Chimica ''E.\ Segr\`e'', Palermo, Italy
\item[$^{61}$] Universit\`a di Roma ``Tor Vergata'', Dipartimento di Fisica, Roma, Italy
\item[$^{62}$] Universit\`a Torino, Dipartimento di Fisica, Torino, Italy
\item[$^{63}$] Benem\'erita Universidad Aut\'onoma de Puebla, Puebla, M\'exico
\item[$^{64}$] Unidad Profesional Interdisciplinaria en Ingenier\'\i{}a y Tecnolog\'\i{}as Avanzadas del Instituto Polit\'ecnico Nacional (UPIITA-IPN), M\'exico, D.F., M\'exico
\item[$^{65}$] Universidad Aut\'onoma de Chiapas, Tuxtla Guti\'errez, Chiapas, M\'exico
\item[$^{66}$] Universidad Michoacana de San Nicol\'as de Hidalgo, Morelia, Michoac\'an, M\'exico
\item[$^{67}$] Universidad Nacional Aut\'onoma de M\'exico, M\'exico, D.F., M\'exico
\item[$^{68}$] Institute of Nuclear Physics PAN, Krakow, Poland
\item[$^{69}$] University of \L{}\'od\'z, Faculty of High-Energy Astrophysics,\L{}\'od\'z, Poland
\item[$^{70}$] Laborat\'orio de Instrumenta\c{c}\~ao e F\'\i{}sica Experimental de Part\'\i{}culas -- LIP and Instituto Superior T\'ecnico -- IST, Universidade de Lisboa -- UL, Lisboa, Portugal
\item[$^{71}$] ``Horia Hulubei'' National Institute for Physics and Nuclear Engineering, Bucharest-Magurele, Romania
\item[$^{72}$] Institute of Space Science, Bucharest-Magurele, Romania
\item[$^{73}$] Center for Astrophysics and Cosmology (CAC), University of Nova Gorica, Nova Gorica, Slovenia
\item[$^{74}$] Experimental Particle Physics Department, J.\ Stefan Institute, Ljubljana, Slovenia
\item[$^{75}$] Universidad de Granada and C.A.F.P.E., Granada, Spain
\item[$^{76}$] Instituto Galego de F\'\i{}sica de Altas Enerx\'\i{}as (IGFAE), Universidade de Santiago de Compostela, Santiago de Compostela, Spain
\item[$^{77}$] IMAPP, Radboud University Nijmegen, Nijmegen, The Netherlands
\item[$^{78}$] Nationaal Instituut voor Kernfysica en Hoge Energie Fysica (NIKHEF), Science Park, Amsterdam, The Netherlands
\item[$^{79}$] Stichting Astronomisch Onderzoek in Nederland (ASTRON), Dwingeloo, The Netherlands
\item[$^{80}$] Universiteit van Amsterdam, Faculty of Science, Amsterdam, The Netherlands
\item[$^{81}$] Case Western Reserve University, Cleveland, OH, USA
\item[$^{82}$] Colorado School of Mines, Golden, CO, USA
\item[$^{83}$] Department of Physics and Astronomy, Lehman College, City University of New York, Bronx, NY, USA
\item[$^{84}$] Michigan Technological University, Houghton, MI, USA
\item[$^{85}$] New York University, New York, NY, USA
\item[$^{86}$] University of Chicago, Enrico Fermi Institute, Chicago, IL, USA
\item[$^{87}$] University of Delaware, Department of Physics and Astronomy, Bartol Research Institute, Newark, DE, USA
\item[] -----
\item[$^{a}$] Max-Planck-Institut f\"ur Radioastronomie, Bonn, Germany
\item[$^{b}$] also at Kapteyn Institute, University of Groningen, Groningen, The Netherlands
\item[$^{c}$] School of Physics and Astronomy, University of Leeds, Leeds, United Kingdom
\item[$^{d}$] Fermi National Accelerator Laboratory, Fermilab, Batavia, IL, USA
\item[$^{e}$] Pennsylvania State University, University Park, PA, USA
\item[$^{f}$] Colorado State University, Fort Collins, CO, USA
\item[$^{g}$] Louisiana State University, Baton Rouge, LA, USA
\item[$^{h}$] now at Graduate School of Science, Osaka Metropolitan University, Osaka, Japan
\item[$^{i}$] Institut universitaire de France (IUF), France
\item[$^{j}$] now at Technische Universit\"at Dortmund and Ruhr-Universit\"at Bochum, Dortmund and Bochum, Germany
\end{description}

\section*{Acknowledgments}

\begin{sloppypar}
The successful installation, commissioning, and operation of the Pierre
Auger Observatory would not have been possible without the strong
commitment and effort from the technical and administrative staff in
Malarg\"ue. We are very grateful to the following agencies and
organizations for financial support:
\end{sloppypar}

\begin{sloppypar}
Argentina -- Comisi\'on Nacional de Energ\'\i{}a At\'omica; Agencia Nacional de
Promoci\'on Cient\'\i{}fica y Tecnol\'ogica (ANPCyT); Consejo Nacional de
Investigaciones Cient\'\i{}ficas y T\'ecnicas (CONICET); Gobierno de la
Provincia de Mendoza; Municipalidad de Malarg\"ue; NDM Holdings and Valle
Las Le\~nas; in gratitude for their continuing cooperation over land
access; Australia -- the Australian Research Council; Belgium -- Fonds
de la Recherche Scientifique (FNRS); Research Foundation Flanders (FWO),
Marie Curie Action of the European Union Grant No.~101107047; Brazil --
Conselho Nacional de Desenvolvimento Cient\'\i{}fico e Tecnol\'ogico (CNPq);
Financiadora de Estudos e Projetos (FINEP); Funda\c{c}\~ao de Amparo \`a
Pesquisa do Estado de Rio de Janeiro (FAPERJ); S\~ao Paulo Research
Foundation (FAPESP) Grants No.~2019/10151-2, No.~2010/07359-6 and
No.~1999/05404-3; Minist\'erio da Ci\^encia, Tecnologia, Inova\c{c}\~oes e
Comunica\c{c}\~oes (MCTIC); Czech Republic -- GACR 24-13049S, CAS LQ100102401,
MEYS LM2023032, CZ.02.1.01/0.0/0.0/16{\textunderscore}013/0001402,
CZ.02.1.01/0.0/0.0/18{\textunderscore}046/0016010 and
CZ.02.1.01/0.0/0.0/17{\textunderscore}049/0008422 and CZ.02.01.01/00/22{\textunderscore}008/0004632;
France -- Centre de Calcul IN2P3/CNRS; Centre National de la Recherche
Scientifique (CNRS); Conseil R\'egional Ile-de-France; D\'epartement
Physique Nucl\'eaire et Corpusculaire (PNC-IN2P3/CNRS); D\'epartement
Sciences de l'Univers (SDU-INSU/CNRS); Institut Lagrange de Paris (ILP)
Grant No.~LABEX ANR-10-LABX-63 within the Investissements d'Avenir
Programme Grant No.~ANR-11-IDEX-0004-02; Germany -- Bundesministerium
f\"ur Bildung und Forschung (BMBF); Deutsche Forschungsgemeinschaft (DFG);
Finanzministerium Baden-W\"urttemberg; Helmholtz Alliance for
Astroparticle Physics (HAP); Helmholtz-Gemeinschaft Deutscher
Forschungszentren (HGF); Ministerium f\"ur Kultur und Wissenschaft des
Landes Nordrhein-Westfalen; Ministerium f\"ur Wissenschaft, Forschung und
Kunst des Landes Baden-W\"urttemberg; Italy -- Istituto Nazionale di
Fisica Nucleare (INFN); Istituto Nazionale di Astrofisica (INAF);
Ministero dell'Universit\`a e della Ricerca (MUR); CETEMPS Center of
Excellence; Ministero degli Affari Esteri (MAE), ICSC Centro Nazionale
di Ricerca in High Performance Computing, Big Data and Quantum
Computing, funded by European Union NextGenerationEU, reference code
CN{\textunderscore}00000013; M\'exico -- Consejo Nacional de Ciencia y Tecnolog\'\i{}a
(CONACYT) No.~167733; Universidad Nacional Aut\'onoma de M\'exico (UNAM);
PAPIIT DGAPA-UNAM; The Netherlands -- Ministry of Education, Culture and
Science; Netherlands Organisation for Scientific Research (NWO); Dutch
national e-infrastructure with the support of SURF Cooperative; Poland
-- Ministry of Education and Science, grants No.~DIR/WK/2018/11 and
2022/WK/12; National Science Centre, grants No.~2016/22/M/ST9/00198,
2016/23/B/ST9/01635, 2020/39/B/ST9/01398, and 2022/45/B/ST9/02163;
Portugal -- Portuguese national funds and FEDER funds within Programa
Operacional Factores de Competitividade through Funda\c{c}\~ao para a Ci\^encia
e a Tecnologia (COMPETE); Romania -- Ministry of Research, Innovation
and Digitization, CNCS-UEFISCDI, contract no.~30N/2023 under Romanian
National Core Program LAPLAS VII, grant no.~PN 23 21 01 02 and project
number PN-III-P1-1.1-TE-2021-0924/TE57/2022, within PNCDI III; Slovenia
-- Slovenian Research Agency, grants P1-0031, P1-0385, I0-0033, N1-0111;
Spain -- Ministerio de Ciencia e Innovaci\'on/Agencia Estatal de
Investigaci\'on (PID2019-105544GB-I00, PID2022-140510NB-I00 and
RYC2019-027017-I), Xunta de Galicia (CIGUS Network of Research Centers,
Consolidaci\'on 2021 GRC GI-2033, ED431C-2021/22 and ED431F-2022/15),
Junta de Andaluc\'\i{}a (SOMM17/6104/UGR and P18-FR-4314), and the European
Union (Marie Sklodowska-Curie 101065027 and ERDF); USA -- Department of
Energy, Contracts No.~DE-AC02-07CH11359, No.~DE-FR02-04ER41300,
No.~DE-FG02-99ER41107 and No.~DE-SC0011689; National Science Foundation,
Grant No.~0450696, and NSF-2013199; The Grainger Foundation; Marie
Curie-IRSES/EPLANET; European Particle Physics Latin American Network;
and UNESCO.
\end{sloppypar}

\end{document}